# Numerical modeling of the friction stir welding process: a literature review

**Diogo Mariano Neto • Pedro Neto**

D. M. Neto, P. Neto

*Department of Mechanical Engineering (CEMUC) - POLO II, University of Coimbra, 3030-788 Coimbra, Portugal*

Tel.: +351 239 790 700

Fax: +351 239 790 701

E-mail: diogo.neto@dem.uc.pt

Corresponding author: D. M. Neto – diogo.neto@dem.uc.pt

**Abstract:**

This survey presents a literature review on friction stir welding (FSW) modeling with a special focus on the heat generation due to the contact conditions between the FSW tool and the workpiece. The physical process is described and the main process parameters that are relevant to its modeling are highlighted. The contact conditions (sliding/sticking) are presented as well as an analytical model that allows estimating the associated heat generation. The modeling of the FSW process requires the knowledge of the heat loss mechanisms, which are discussed mainly considering the more commonly adopted formulations. Different approaches that have been used to investigate the material flow are presented and their advantages/drawbacks are discussed. A reliable FSW process modeling depends on the fine tuning of some process and material parameters. Usually, these parameters are achieved with base on experimental data. The numerical modeling of the FSW process can help to achieve such parameters with less effort and with economic advantages.

## 1. INTRODUCTION

### 1.1. Friction Stir Welding Process

Friction stir welding (FSW) is a novel solid state joining process patented in 1991 by The Welding Institute, Cambridge, UK [1]. One of the main advantages of FSW over the conventional fusion joining techniques is that no melting occurs. Thus, the FSW process is performed at much lower temperatures than the conventional welding. At the same time, FSW allows to avoid many of the environmental and safety issues associated with conventional welding methods [2]. In FSW the parts to weld are joined by forcing a rotating tool to penetrate into the joint and moving across the entire joint. Resuming, the solid-state joining process is promoted by the movement of a non-consumable tool (FSW tool) through the welding joint.

FSW consists mainly in three phases, in which each one can be described as a time period where the welding tool and the workpiece are moved relative to each other. In the first phase, the rotating tool is vertically displaced into the joint line (plunge period). This period is followed by the dwell period in which the tool is held steady relative to the workpiece but still rotating. Owing to the velocity difference between the rotating tool and the stationary workpiece, the mechanical interaction produces heat by means of frictional work and material plastic deformation. This heat is dissipated into the neighboring material, promoting an increase of temperature and consequent material softening. After these two initial phases the welding operation can be initiated by moving either the tool or the workpiece relative to each other along the joint line. Fig. 1 illustrates a schematic representation of the FSW setup [3].

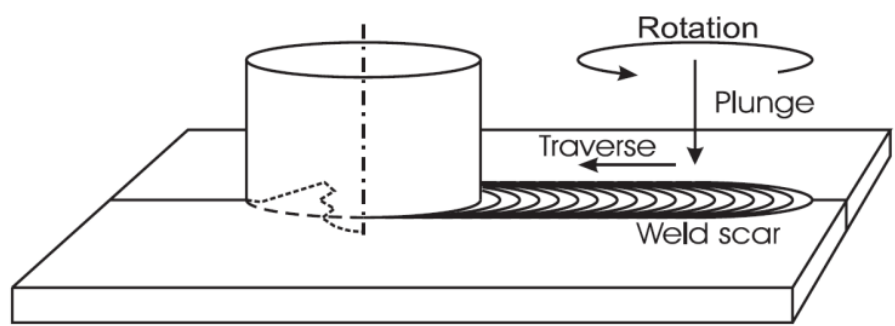

Fig. 1 Friction stir welding setup [3]

The FSW tool consists of a rotating probe (also called pin) connected to a shoulder piece, as shown in Fig. 2. During the welding operation, the tool is moved along the butting surfaces of the two rigidly clamped plates (workpiece), which are normally placed on a backing plate. The vertical displacement of the tool is controlled to guarantee that the shoulder keeps contact with the top surface of the workpiece. The heat generated by the friction effect and plastic deformation softens the material being welded. A severe plastic deformation and flow of plasticized metal occurs when the tool is translated along the welding direction. In this way, the material is transported from the front of the tool to the trailing edge (where it is forged into a joint) [4].

The half-plate in which the direction of the tool rotation is the same as the welding direction is called the advancing side, while the other is designated as retreating side. This difference can lead to asymmetry in heat transfer, material flow and in the mechanical properties of the weld.

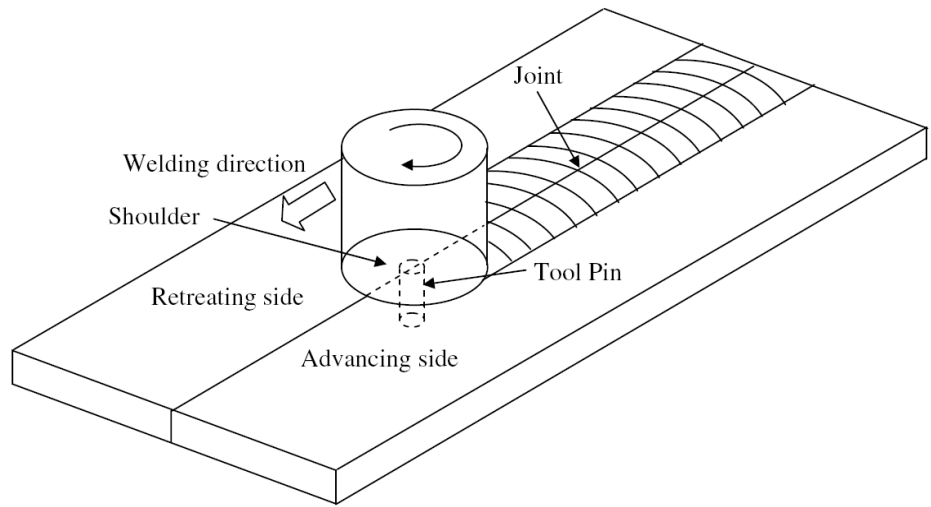

Fig. 2 Schematic illustration of the FSW process [4]

### 1.1.1. Process Parameters

The welding traverse speed ($v_{trans}$), the tool rotational speed ($\omega$), the downward force ($F$), the tilt angle of the tool and the tool design are the main variables usually used to control the FSW process [4]. The rotation of the tool results in stirring of material around the tool probe while the translation of the tool moves the stirred material from the front to the back of the probe. Axial pressure on the tool also affects the quality of the weld. It means that very high pressures lead to overheating and thinning of the joint, whereas very low pressures lead to insufficient heating and voids. The tilt angle of the tool, measured with respect to the workpiece surface, is also an important parameter, especially to help producing welds with "smooth" tool shoulders [5].

As mentioned before, tool design influences heat generation, plastic flow, the power required to perform FSW and the uniformity of the welded joint. Generally, two tool surfaces are needed to perform the heating and joining processes in FSW. The shoulder surface is the area where the majority of the heat by friction is generated. This is valid for relatively thin plates, otherwise the probe surface is the area where the majority of the heat is generated. Fig. 3 presents a schematic example of an FSW tool with conical shoulder and threaded probe. In this case, the conical tool shoulder helps to establish a pressure under the shoulder, but also operates as an escape volume for the material displaced by the probe due to the plunge action. As the probe tip must not penetrate the workpiece or damage the backing plate, in all tool designs the probe height is limited by the workpiece thickness [3].

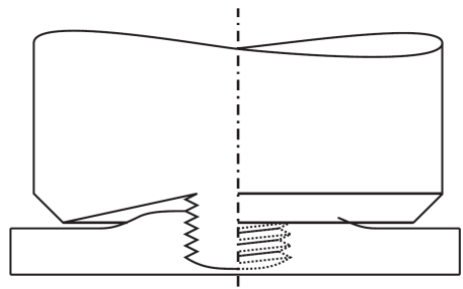

Fig. 3 FSW tool with a conical shoulder and threaded probe [3]

### 1.1.2. Weld Microstructure

FSW involves complex interactions between simultaneous thermomechanical processes. These interactions affect the heating and cooling rates, plastic deformation and flow, dynamic recrystallization phenomena and the mechanical integrity of the joint [4]. The thermomechanical process involved under the tool results in different microstructural regions (see Fig. 4). Some microstructural regions are common to all forms of welding, while others are exclusive of FSW [5].

- The stir zone (also called nugget) is a region of deeply deformed material that corresponds approximately to the location of the probe during welding. The grains within the nugget are often an order of magnitude smaller than the grains in the base material.
- The thermomechanically affected zone (TMAZ) occurs on either side of the stir zone. The strain and temperature levels attained are lower and the effect of welding on the material microstructure is negligible.
- The heat affected zone (HAZ) is common to all welding processes. This region is subjected to a thermal cycle but it is not deformed during welding.

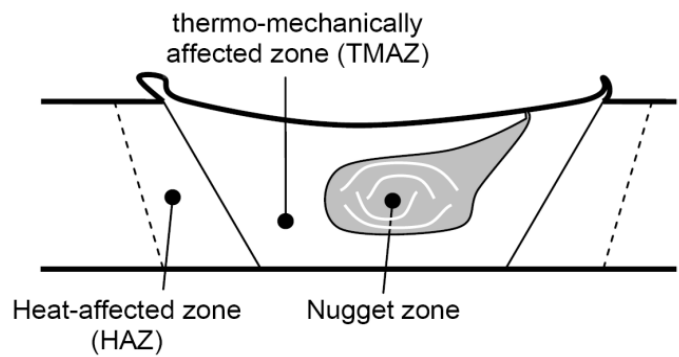

Fig. 4 Different microstructural regions in a transverse cross section of FSW [5]

**1.2. Numerical Modeling**

Several aspects of the FSW process are still poorly understood and require further study. Many experimental investigations have already been conducted to adjust input FSW parameters (tool speed, feed rate and tool depth), contrary to numerical investigations, which have been scarcely used for these purposes. Computational tools could be helpful to better understand and visualize the influence of input parameters on FSW process. Visualization and analysis of the material flow, temperature field, stresses and strains involved during the FSW process can be easily obtained using simulation results than using experimental ones. Therefore, in order to attain the best weld properties, simulations can help to adjust and optimize the process parameters and tool design [5].

One of the main research topics in FSW is the evaluation of the temperature field [6]. Although the temperatures involved in the process are lower than the melting points of the weld materials, they are high enough to promote phase transformations. Thus, it is very important to know the time-temperature history of the welds. Usually, FSW temperature is measured using thermocouples [7-8]. However, the process of measuring temperature variations in the nugget zone using the technique mentioned above is a very difficult task. Numerical methods can be very efficient and convenient for this study and in fact, along the last few years, they have been used in the field of FSW [9]. Riahi and Nazari present numerical results indicating that the high gradient in temperature (for an aluminum alloy) is in the region under the shoulder [10].

In the process modeling, it is essential to keep the goals of the model in view and at the same time it is also important to adopt an appropriate level of complexity. In this sense, both analytical and numerical methods have a role to play [11]. Usually, two types of process modeling techniques are adopted: fluid dynamics (simulation of material flow and temperature distribution) and solid mechanics (simulation of temperature distribution, stress and strain). Both solid and fluid modeling techniques involve non-linear phenomena belonging to the three classic types: geometric, material or contact nonlinearity.

The simulation of material flow during FSW has been modeled using computational fluid dynamics (CFD) formulations. In this scenario, the material is analyzed as a viscous fluid flowing across an Eulerian mesh and interacting with a rotating tool [12]. Other authors have also used a CFD approach to develop a global thermal model in which the heat flow model includes parameters related with the shear material and friction phenomenon [13]. One of the major disadvantages of CFD models has to do with the definition of the material properties (residual stresses cannot be predicted) [7].

Solid mechanics models require the use of Lagrangian formulation due to the high deformation levels. However, the high gradient values of the state variables near to the probe and the thermomechanical coupling imply a large number of degrees of freedom in FSW modeling, which is costly in terms of CPU time [14]. Recent research demonstrated that the computational time can be reduced by recurring to high performance computing (HPC) techniques [15].

Nevertheless, in order to face the long computational times associated to the simulation of the FSW process, the adaptive arbitrary Lagrangian Eulerian (ALE) formulation has been implemented by some authors [16-17]. Van der Stelt et al. use an ALE formulation to simulate the material flow around the pin during FSW process [16]. These models of the process can predict the role played by the tool plunge depth on the formation of flashes, voids or tunnel defects, and the influence of threads on the material flow, temperature field and welding forces [14]. Lagrangian, Eulerian and ALE approaches have been used to numerically simulate the FSW process, using software such as FORGE3 and THERCAST [18], ABAQUS [10], DiekA [16], WELDSIM [19] and SAMCEF [20].

## 2. HEAT GENERATION

The heat generated during the welding process is equivalent to the power input introduced into the weld by the tool minus some losses due to microstructural effects [21]. The peripheral speed of the shoulder and probe is much higher than the translational speed (the tool rotates at high speeds). FSW primarily uses viscous dissipation in the workpiece material, driven by high shear stresses at the tool/workpiece interface. Therefore, the heat generation modeling requires some representation of the behaviour of the contact interface, together with the viscous dissipation behaviour of the material. However, the boundary conditions in FSW are complex to define. Material at the interface may either stick to the tool (it has the same local velocity as the tool) or it may slip (the velocity may be lower) [11]. An analytical model for heat generation in FSW based on different assumptions in terms of contact condition between the rotating tool surface and the weld piece was developed by Schmidt et al. [3]. This model will be discussed in the following sections.

### 2.1. Contact Condition

When modeling the FSW process, the contact condition is a critical part of the numerical model [22]. Usually, the Coulomb friction law is applied to describe the shear forces between the tool surface and the workpiece. In general, the law estimates the contact shear stress as:

$$\tau_{friction} = \mu p \qquad (1)$$

where $\mu$ is the friction coefficient and $p$ is the contact pressure. Analyzing the contact condition of two infinitesimal surface segments in contact, Coulomb's law predicts the mutual motion between the two segments (whether they stick or slide). The normal interpretation of Coulomb's law is based on rigid contact pairs, without taking into account the internal stress. However, this is not sufficiently representative for the FSW process. Thus, three different contact states were developed at the tool/workpiece interface, and they can be categorized according to the definition presented by Schmidt et al. [3].

### 2.1.1. Sliding Condition

If the contact shear stress is smaller than the internal matrix (material to be welded) yield shear stress, the matrix segment volume shears slightly to a stationary elastic deformation (sliding condition).

### 2.1.2. Sticking Condition

When the friction shear stress exceeds the yield shear stress of the underlying matrix, the matrix surface will stick to the moving tool surface segment. In this case, the matrix segment will accelerate along the tool surface (receiving the tool velocity), until the equilibrium state is established between the contact shear stress and the internal matrix shear stress. At this point, the stationary full sticking condition is fulfilled. In conventional Coulomb's friction law terms, the static friction coefficient relates the reactive stresses between the surfaces.

### 2.1.3. Partial Sliding/ Sticking Condition

The last possible state between the sticking and sliding condition is a mixed state of both. In this case, the matrix segment accelerates to a velocity less than the tool surface velocity. The equilibrium is established when the contact shear stress equals the internal yield shear stress due to a quasi-stationary plastic deformation rate (partial sliding/sticking condition). In resume, the sliding condition promotes heat generation by means of friction and the sticking condition promotes heat generation by means of plastic deformation. In practice, we have these two conditions together (partial sliding/sticking condition).

### 2.1.4. Contact State Variable

It is convenient to define a contact state variable $\delta$, which relates the velocity of the contact workpiece surface with the velocity of the tool surface. This parameter is a dimensionless slip rate defined by Schmidt et al. [3] as:

$$\delta = \frac{v_{workpiece}}{v_{tool}} = 1 - \frac{\dot{\gamma}}{v_{tool}} \qquad (2)$$

$$\dot{\gamma} = v_{tool} - v_{workpiece} \qquad (3)$$

where $v_{tool}$ is the velocity of the tool calculated from $\omega r$ (being $\omega$ the angular velocity and $r$ the radius), $v_{workpiece}$ is the local velocity of the matrix point at the tool/workpiece contact interface and $\dot{\gamma}$ is the slip rate. Furthermore, the assumption that the welding transverse speed does not influence the slip rate and/or the deformation rate, results in that all workpiece velocities can be considered tangential to the rotation axis. It is then possible to define $\delta$ as:

$$\delta = \frac{\omega_{workpiece}}{\omega_{tool}} \qquad (4)$$

where $\omega_{workpiece}$ is the angular rotation speed of the contact matrix layer and $\omega_{tool}$ is the angular rotation speed of the tool. Ulysse uses this relationship to prescribe a slip boundary condition in his CFD models of the material flow in FSW [23]. The relationship between the different contact conditions is summarized in Table I.

Table I: Definition of contact condition, velocity/shear relationship and state variable ( $\dot{\varepsilon}$ strain rate ) [24]

| Contact condition | Matrix velocity | Tool velocity | Contact shear stress | State variable |
|---|---|---|---|---|
| Sticking | $v_{matrix} = v_{tool}$ | $v_{tool} = \omega r$ | $\tau_{contact} = \tau_{yield}(\uparrow \dot{\varepsilon})$ | $\delta = 1$ |
| Sticking/sliding | $v_{matrix} < v_{tool}$ | $v_{tool} = \omega r$ | $\tau_{yield}(\downarrow \dot{\varepsilon}) < \tau_{contact} < \tau_{yield}(\uparrow \dot{\varepsilon})$ | $0 < \delta < 1$ |
| Sliding | $v_{matrix} = 0$ | $v_{tool} = \omega r$ | $\tau_{contact} < \tau_{yield}$ | $\delta = 0$ |

## 2.2. Analytical Estimation of Heat Generation

During the FSW process, heat is generated close to the contact surfaces, which can have complex geometries according to the tool geometry. However, for the analytical model, it is assumed a simplified tool design with a conical or horizontal shoulder surface, a vertical cylindrical probe surface and an horizontal probe tip surface. The conical shoulder surface is characterized by the cone angle $\alpha$, which in the case of a flat shoulder assumes the value zero. The simplified tool design is presented in Fig. 5, where $R_{shoulder}$ is the radius of the shoulder, and $R_{probe}$ and $H_{probe}$ is the probe radius and height, respectively. Fig. 5 also represents the heat generated under the tool shoulder $Q_1$, the tool probe side $Q_2$, and at the tool probe tip $Q_3$. In this way, the total heat generation can be calculated $Q_{total} = Q_1 + Q_2 + Q_3$. The heat generated in each contact surface can then be computed [24]:

$$dQ = \omega dM = \omega r dF = \omega r \tau_{contact} dA \qquad (5)$$

where $M$ is the moment, $F$ is the force, $A$ is the contact area and $r$ is the cylindrical coordinate.

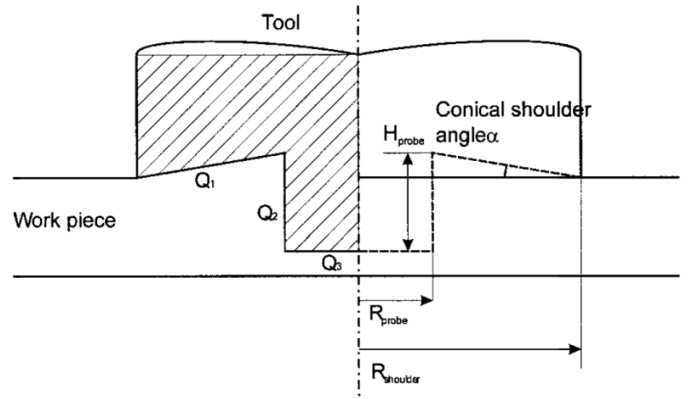

Fig. 5 Heat generation contributions represented in a simplified FSW tool [17]

### 2.2.1. Heat Generation

There follows heat generation derivations which are analytical estimations of the heat generated at the contact interface between a rotating FSW tool and a stationary weld piece matrix. The mechanical power due to the traverse movement is not considered, as this quantity is negligible compared to the rotational power. A given surface of the tool in contact with the matrix is characterized by its position and orientation relative to the rotation axis of the tool, as shown in Fig. 6.

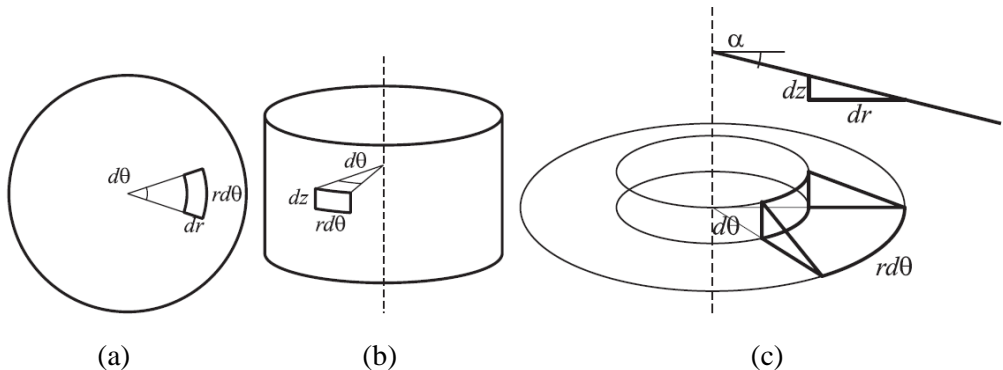

(a)          (b)          (c)

Fig. 6 Surface orientations and infinitesimal segment areas: (a) Horizontal; (b) Vertical; (c) Conical/tilted [3]

#### 2.2.1.1. Heat Generation from the Shoulder

The shoulder surface of a modern FSW tool is in most cases concave or conically shaped. Previous analytical expressions for heat generation include a flat circular shoulder, in some cases omitting the contribution from the probe [25]. Schmidt et al. extends the previous expressions so that the conical shoulder and cylindrical probe surfaces are included in the analytical expressions [24]. This analytical model for the heat generation phenomena does not include non-uniform pressure distribution, strain rate dependent yield shear stresses and the material flow driven by threads or flutes. Integration over the shoulder area from $R_{probe}$ to $R_{shoulder}$ using equation (5) gives the shoulder heat generation:

$$Q_1 = \int_0^{2\pi} \int_{R_{probe}}^{R_{shoulder}} \omega \tau_{contact} r^2 (1+\tan\alpha) dr d\theta$$
$$= \frac{3}{2}\pi\omega\tau_{contact}(R_{shoulder}^3 - R_{probe}^3)(1+\tan\alpha) \tag{6}$$

#### 2.2.1.2. Heat Generation from the Probe

The heat generated at the probe has two contributions: $Q_2$ from the side surface and $Q_3$ from the tip surface. Integrating over the probe side area:

$$Q_2 = \int_0^{2\pi} \int_0^{H_{probe}} \omega \tau_{contact} R_{probe}^2 dz d\theta = 2\pi\omega\tau_{contact} R_{probe}^2 H_{probe} \tag{7}$$

and integrating the heat flux based over the probe tip surface, assuming a flat tip, we have that:

$$Q_3 = \int_0^{2\pi} \int_0^{R_{probe}} \omega \tau_{contact} r^2 dr d\theta = \frac{2}{3}\pi\omega\tau_{contact} R_{probe}^3 \tag{8}$$

The three contributions are combined to get the total heat generation estimate:

$$Q_{total} = Q_1 + Q_2 + Q_3$$
$$= \frac{2}{3}\pi\omega\tau_{contact}((R_{shoulder}^3 - R_{probe}^3)(1+\tan\alpha) + R_{probe}^3 + 3R_{probe}^2 H_{probe}) \tag{9}$$

In the case of a flat shoulder, the heat generation expression simplifies to:

$$Q_{total} = \frac{2}{3}\pi\omega\tau_{contact}(R_{shoulder}^3 + 3R_{probe}^2 H_{probe}) \tag{10}$$

This last expression correlates with the results obtained by Khandkar et al. [26].

### 2.2.2. Influence of contact status: sticking and sliding

Equation (9) is based on the general assumption of a constant contact shear stress, but the mechanisms behind the contact shear stress vary, depending on whether the material verifies the sliding or sticking condition. If the sticking interface condition is assumed, the matrix closest to the tool surface sticks to it. The layer between the stationary material points and the material moving with the tool has to accommodate the velocity difference by shearing. The contact shear stress is then:

$$\tau_{contact} = \tau_{yield} = \frac{\sigma_{yield}}{\sqrt{3}} \tag{11}$$

This gives a modified expression of (9), assuming the sticking condition:

$$Q_{total,sticking} = \frac{2}{3}\pi\omega\frac{\sigma_{yield}}{\sqrt{3}}((R_{shoulder}^3 - R_{probe}^3)(1+\tan\alpha) + R_{probe}^3 + 3R_{probe}^2 H_{probe}) \tag{12}$$

Assuming a friction interface condition, where the tool surface and the weld material are sliding against each other, the choice of Coulomb's friction law to describe the shear stress estimates the critical friction stress necessary for a sliding condition:

$$\tau_{contact} = \tau_{friction} = \mu p \tag{13}$$

Thus, for the sliding condition, the total heat generation is given by:

$$Q_{total,sliding} = \frac{2}{3}\pi\omega\mu p((R_{shoulder}^3 - R_{probe}^3)(1+\tan\alpha) + R_{probe}^3 + 3R_{probe}^2 H_{probe}) \tag{14}$$

The analytical solution for the heat generation under the partial sliding/sticking condition is simply a combination of the two solutions, respectively, with a kind of weighting function. From the partial sliding/sticking condition follows that the slip rate between the surfaces is a fraction of $\omega r$, lowering the heat generation from sliding friction. This is counterbalanced by the additional plastic dissipation due to material deformation. This enables a linear combination of the expressions for sliding and sticking:

$$\begin{aligned} Q_{total} &= \delta Q_{total,sticking} + (1-\delta)Q_{total,sliding} = \\ &= \frac{2}{3}\pi\omega(\delta\tau_{yield} + (1-\delta)\mu p)((R_{shoulder}^3 - R_{probe}^3)(1+\tan\alpha) + R_{probe}^3 + 3R_{probe}^2 H_{probe}) \end{aligned} \tag{15}$$

where $\delta$ is the contact state variable (dimensionless slip rate), $\tau_{yield}$ is the material yield shear stress at welding temperature, $\omega$ is the angular rotation speed and $\alpha$ is the cone angle. This expression (15) can be used to estimate the heat generation for $0 \leq \delta \leq 1$, corresponding to sliding when $\delta = 0$, sticking when $\delta = 1$ and partial sliding/sticking when $0 < \delta < 1$.

### 2.2.3. Heat Generation Ratios

Based on the geometry of the tool and independently from the contact condition, the ratios of heat generation are as follows:

$$f_{shoulder} = \frac{Q_1}{Q_{total}} = \frac{(R_{shoulder}^3 - R_{probe}^3)(1+\tan\alpha)}{(R_{shoulder}^3 - R_{probe}^3)(1+\tan\alpha) + R_{probe}^3 + 3R_{probe}^2 H_{probe}} = 0.86 \tag{16}$$

$$f_{probe-side} = \frac{Q_2}{Q_{total}} = \frac{3R_{probe}^2 H_{probe}}{(R_{shoulder}^3 - R_{probe}^3)(1+\tan\alpha) + R_{probe}^3 + 3R_{probe}^2 H_{probe}} = 0.11 \tag{17}$$

$$f_{probe-tip} = \frac{Q_3}{Q_{total}} = \frac{R_{probe}^3}{(R_{shoulder}^3 - R_{probe}^3)(1+\tan\alpha) + R_{probe}^3 + 3R_{probe}^2 H_{probe}} = 0.03 \tag{18}$$

where the considered tool dimensions are $R_{shoulder} = 9$ mm, $R_{probe} = 3$ mm, $H_{probe} = 4$ mm and $\alpha = 10°$. This indicates that the heat generation from the probe is negligible for a thin plate, but, it is typically 10% or more for a thick plate [11]. Fig. 7 presents the evolution of the heat generation ratios of the shoulder and probe as a function of the probe radius. Also, in Fig. 7 the influence of the $R_{shoulder}/R_{probe}$ ratio in the heat generation ratio is highlighted.

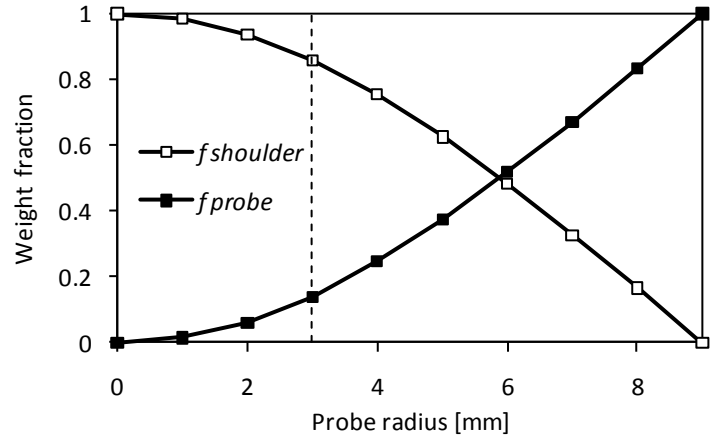

Fig. 7 Heat fraction generated by the shoulder and probe ($R_{shoulder}=9$ mm, $H_{probe}=4$ mm and $\alpha=10°$)

The analytical heat generation estimate correlates with the experimental heat generation, assuming either a sliding or a sticking condition. In order to estimate the experimental heat generation for the sliding condition, a friction coefficient that lies in the reasonable range of known metal to metal contact values is used. Assuming the sticking condition, a yield shear stress, which is descriptive of the weld piece material at elevated temperatures, is used to correlate the values [24].

## 2.3. Heat Generation Mechanism

It is important to mention that it is not even clear what is the nature of the tool interface contact condition, particularly for the shoulder interface. Frigaard et al. developed a numerical model for FSW based on the finite difference method [27]. They assumed that heat is generated at the tool shoulder due to frictional heating and the friction coefficient is adjusted so that the calculated peak temperature did not exceed the melting temperature. Zahedul et al. concluded that a purely friction heating model is probability not adequate (a low value for the friction coefficient was used) [28]. The high temperature values measured by Tang et al. near the pin suggest that heat is generated mainly through plastic deformation during the FSW process [29]. Colegrove et al. assume that the material is completely sticking to the tool [30]. The heating volumetric region where plastic dissipation occurs is the thermomechanically heat affected zone (TMAZ). The corresponding volume heat sources are equal to:

$$q_v = \beta \dot{\varepsilon}_{ij}^p \sigma_{ij} \qquad i,j=1,2,3 \tag{19}$$

where $\dot{\varepsilon}_{ij}^p$ and $\sigma_{ij}$ are the components of the plastic strain rate tensor and the Cauchy tensor, respectively. Also in (19), $\beta$ is a parameter, known as the Taylor-Quinney coefficient, ranging typically between 0.8 and 0.99 [31].

### 2.3.1. Surface and Volume Heat Contributions

The heat input can be divided into surface and volume heat contributions due to frictional or viscous (plastic dissipation) heating, respectively. Simar et al. introduce a parameter ($\gamma$) that exposes the relative importance of both contributions [32]:

$$Q_v = \gamma Q \tag{20}$$

$$Q_s = (1-\gamma)Q \tag{21}$$

where $Q_v$ is the volume heat contribution and $Q_s$ is the total tool surface heat contribution. For thermal computational models which take into account the material fluid flow, Simar et al. concluded that a value of $\gamma = 1$ produces a best agreement with experimental thermal data [32].

### 2.4. Heat Input Estimation using the Torque

Modern FSW equipment usually outputs the working torque as well as the working angular velocity. The power spent in the translation movement, which is approximately 1% of the total value, is typically neglected in the total heat input estimative [11, 30]. Therefore, the power introduced by the tool (input power $P$) can be obtained experimentally from the weld moment and angular rotation speed [21, 32]:

$$P = M\omega + \underbrace{F_{trans} v_{trans}}_{\text{negligible}} \tag{22}$$

where $\omega$ is the tool rotational speed (rad/s), $M$ is the measured torque (N.m), $F_{trans}$ is the traverse force (N) and $v_{trans}$ (m/s) is the traverse velocity. Therefore, the heat input near the interface is given by:

$$Q = P\eta \tag{23}$$

where $\eta$ is the fraction of power generated by the tool that is directly converted into heat in the weld material. Nandan et al. refer to this as the power efficiency factor [33]. This value is usually high, between 0.9 and 1.0, and it is calculated based on the heat loss into the tool, as will be show in the next section.

## 3. HEAT DISSIPATION

Heat generation and heat dissipation must be adjusted and balanced to obtain an agreement with experimental temperature values [34]. As mentioned before, the heat in FSW is generated by the frictional effect and by plastic deformation associated with material stirring. The heat is dissipated into the workpiece leading to the TMAZ and the HAZ, depending on the thermal conductivity coefficient of the base material. The heat loss occurs by means of conduction to the tool and the backing plate, and also by means of convective heat loss to the surrounding atmosphere. The heat lost through convection/radiative is considered negligible [33].

### 3.1. Heat Loss into the Tool

Only a small fraction of the heat is lost into the tool itself. This value may be estimated from a simple heat flow model for the tool. Measuring the temperature at two locations along the tool axis, allows a simple evaluation of the heat losses into the tool. The value of the heat loss into the tool has been studied using this approach, leading to similar conclusions. After modeling the temperature distributions in the tool and comparing it with experimental results, various authors conclude that the heat loss is about 5% [32, 24].

### 3.2. Heat Loss by the Top Surface of the Workpiece

The boundary condition for heat exchange between the top surface of the workpiece and the surroundings, beyond the shoulder, involves considering both the convective and the radiative heat transfer, which can be estimated using the following differential equation [33]:

$$-k\frac{\partial T}{\partial z}\bigg|_{top} = \sigma\varepsilon(T^4 - T_a^4) + h(T - T_a) \tag{24}$$

where $\sigma$ is the Stefan–Boltzmann constant, $\varepsilon$ is the emissivity, $T_a$ is the ambient temperature and $h$ is the heat transfer coefficient at the top surface.

### 3.3. Heat Loss by the Bottom Surface of the Workpiece

Most of the FSW process heat is dissipated through the backing plate due to the contact with the clamps. The heat loss through the contact interface between the bottom of the workpiece and the backing plate has been introduced in numerical models using different approaches [8]. In fact, the contact conditions between the workpiece and the backing plate must be carefully described at the moment of the modeling process. Thus various options can be considered:
- No backing plate. The lower surface of the workpiece is assumed to be adiabatic;
- Perfect contact between workpiece and backing plate;

- Perfect contact under the tool region only. This option is suggested by experimental observations: the high pressures under the tool lead to a visible indentation of the upper surface of the blanking plate along a width approximately equal to the diameter of the tool shoulder (Fig. 8);
- Introduction of a value for the convection coefficient between the workpiece and the backing plate.

Ulysse did not include the backing plate in the model, using the assumption of simply adiabatic conditions at the workpiece/backing interface [23]. A reasonable agreement between predicted and measured temperatures was attained, although measured temperatures tended to be consistently over-predicted by the model. Other authors consider the presence of a backing plate in the model and simulate the contact condition between the workpiece and the backing plate. Colegrove et al. proposed a contact conductance of 1000 $Wm^{-2}K^{-1}$ between the workpiece and the backing plate, except under the tool region where a perfect contact is modeled [35].

The majority of dissipated heat flows from the workpiece to the backing plate at the interface under the tool. Owing to the applied pressure, the conductance gap in this location is smaller than the conductance gap to the surrounding areas, and by this way locally maximizing the heat flow. The use of a backing spar, in opposition to a fully backing plate, reduces the number of equations to be solved and shortens the computer processing time, while still capturing the essential nature of heat flow between the workpiece and backing plate [2] (Fig. 8). The width of the backing spar is usually equal to the tool diameter, and the height varies within the thicknesses range of the backing plate. Khandkar et al. use a 12 mm backing plate [26], Hamilton et al. assume 25.4 mm [2], while Colegrove et al. adopt a 60 mm backing plate [13]. It can be concluded that the larger the thickness of the backing plate, the greater the heat dissipation.

Zahedul et al. propose a value for the convection coefficient between the workpiece and the backing plate by comparing the results of their 3D finite element models with the experimental results [28]. They compare four different bottom convection coefficients and conclude that a value too high for this coefficient leads to an underestimating of the maximum temperature.

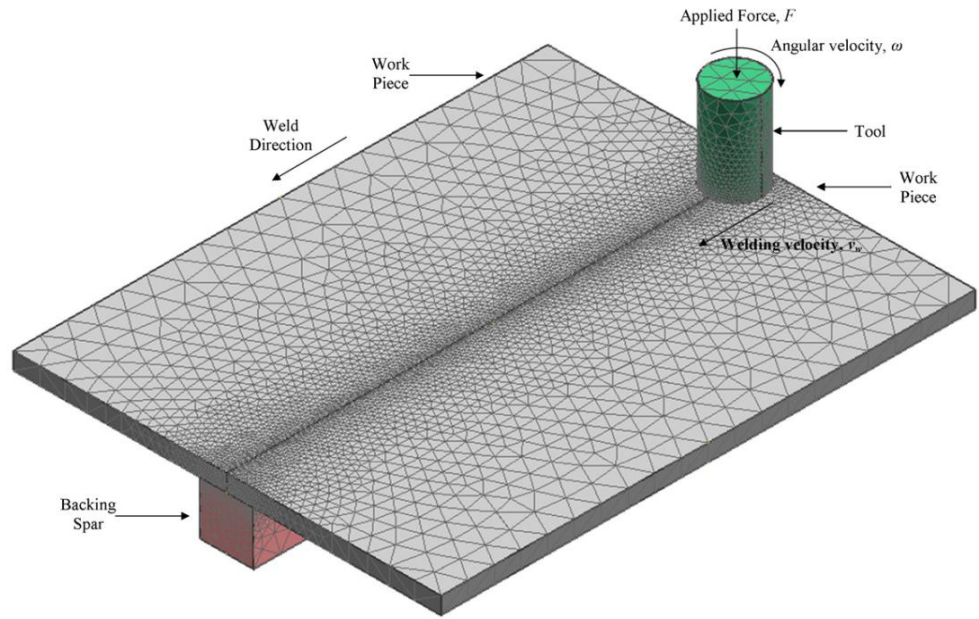

Fig. 8 Employing a backing spar to model the contact condition between workpiece and backing plate [2]

## 4. METAL FLOW

Material flow during FSW is quite complex, it depends on the tool geometry, process parameters and material to be welded. It is of practical importance to understand the material flow characteristics for optimal tool design and to obtain high structural efficiency welds [36]. Modeling of the metal flow in FSW is a challenging problem, but this is a fundamental task to understanding and developing the process. Flow models should be able to simultaneously capture the thermal and mechanical aspects of a given problem in adequate detail to address the following topics:

- Flow visualization, including the flow of dissimilar metals;
- Evaluation of the heat flow that governs the temperature field;
- Tool design to optimize tool profiling for different materials and thicknesses;
- Susceptibility to formation of defects.

The material flow around the probe is one of the main parameters, determinant for the success of FSW [36]. Some studies show that the flow occurs predominantly in the plate plane. Hence, various authors have first analyzed the 2-D flow around the probe at midthickness rather than the full 3-D flow. This produces significant benefits in computational efficiency [37].

Schneider et al. have based their physical model of the metal flow in the FSW process in terms of the kinematics describing the metal motion [38]. This approach has been followed by other authors. Fig. 9 illustrates the decomposition of the FSW process into three incompressible flow fields, combined to create two distinct currents. In this model, a rigid body rotation field imposed by the axial rotation of the probe tool is modified by a superimposed ring vortex field encircling the probe imposed by the pitch of the weld probe threads. These two flow fields, bound by a shear zone, are uniformly translated down the length of the weld panel [36].

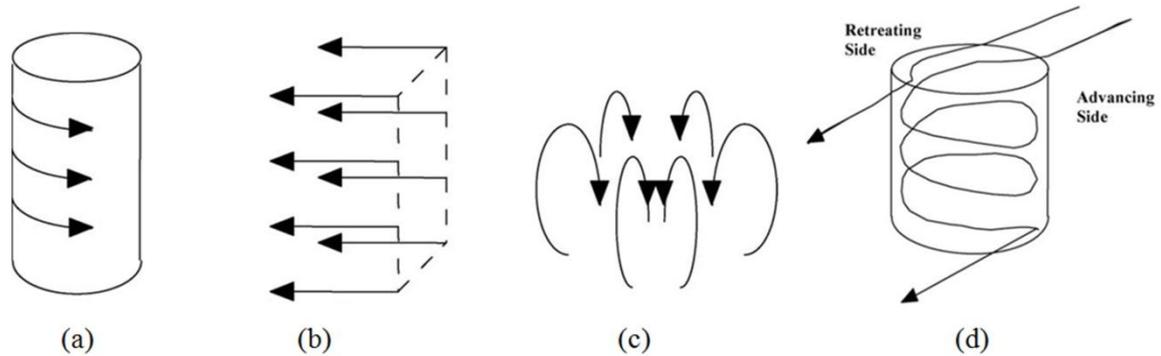

Fig. 9 Schematic representation of the three incompressible flow fields of the friction stir weld: (a) rotation; (b) translation; (c) ring vortex; (d) summation of three flow fields [39]

A number of approaches have been used to visualize material flow pattern in FSW, using a tracer technique by marker or through the welding of dissimilar alloys. In addition, some computational methods including CFD and finite element analysis (FEA) have been also used to model the material flow [36].

### 4.1. Numerical Flow Modeling

Numerical FSW flow modeling can be based on analyses and techniques used for other processes, such as friction welding, extrusion, machining, forging, rolling and ballistic impact [36]. As for heat flow analyses, numerical flow models can use either an Eulerian or Lagrangian formulation for the mesh, other solution can be the combination of both (hybrid solution and Lagrangian-Eulerian).

The CFD analysis of FSW ranges from 2-D flow around a cylindrical pin to full 3-D analysis of flow around a profiled pin [30]. One consequence of using CFD analysis in relation to solid mechanic models is that some mechanical effects are excluded from the scope of the analysis, for example, studying the effect of varying the downforce. These models cannot predict absolute forces because elasticity is neglected. Also, since that for deforming material it is necessary to fill the available space between the solid boundaries, free surfaces also present difficulties in CFD. One difficulty in the numerical analysis is the steep gradient in flow velocity near the tool. In order to solve this problem, most analyses divide the mesh into zones, as shown in Fig. 10. The flow near the tool is predominantly rotational, thus the mesh in this region rotates with the tool. The rotating zone is made large enough to contain the entire deformation zone and the mesh size is much finer in that zone [30].

A 3-D elastic-plastic finite element analysis, using an ALE formulation, provides results with an interesting physical insight. However, they present very long computation times, making it unlikely to be used routinely as a design tool [36]. Note that 3-D analysis is able to handle some of the process complexities: a concave shoulder, tool tilt, and threaded pin profiles.

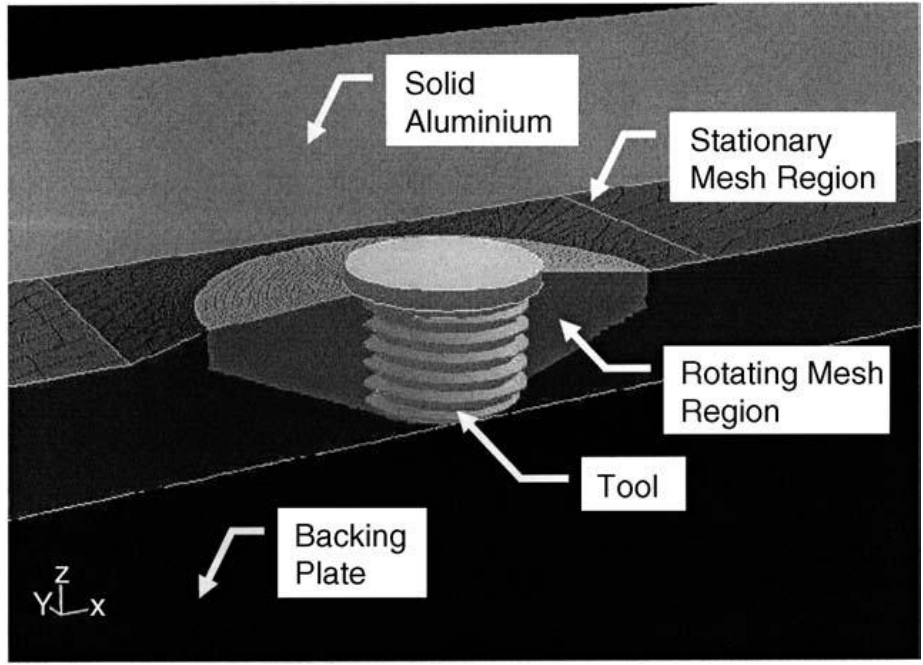

Fig. 10 Mesh definition for computational fluid dynamics analysis of friction stir welding [30]

### 4.1.1. Material Constitutive Behaviour for Flow Modeling

The most common approach to model steady-state hot flow stress is the Sellars-Tegart law, combining the dependence on temperature $T$ and strain rate $\dot{\varepsilon}$ *via* the Zener-Hollomon parameter:

$$Z = \dot{\varepsilon} \exp\left(\frac{Q}{RT}\right) = A(\sinh\alpha\sigma)^n \tag{25}$$

where $Q$ is an effective activation energy, $R$ is the gas constant and $\alpha$, $A$, and $n$ are material parameters. Other authors have used an alternative constitutive response developed by Johnson and Cook for modeling ballistic impacts [40]:

$$\sigma_y = \left[A + B(\overline{\varepsilon}^{pl})^n\right]\left(1 + C\ln\frac{\dot{\overline{\varepsilon}}^{pl}}{\dot{\varepsilon}_0}\right)\left(1 - \left(\frac{T - T_{ref}}{T_{melt} - T_{ref}}\right)^m\right) \tag{26}$$

where $\sigma_y$ is the yield stress, $\overline{\varepsilon}^{pl}$ the effective plastic strain, $\dot{\overline{\varepsilon}}^{pl}$ the effective plastic strain rate, $\dot{\varepsilon}_0$ the normalizing strain rate, and $A$, $B$, $C$, $n$, $T_{melt}$, $T_{ref}$ and $m$ are material/test parameters. Mishra and Ma reported that the general flow pattern predicted is rather insensitive to the constitutive law due to the inherent kinematic constraint of the process [36]. However, the heat generation, temperature, and flow stress near the tool and the loading on the tool will depend closely on the material law.

## 5. NUMERICAL SIMULATION OF FSW

A correct model of the FSW process should avoid any unnecessary assumptions. A list of requirements for a FSW analysis code includes the following:
- Rotational boundary condition;
- Frictional contact algorithms;
- Support very high levels of deformation;
- Elastic-Plastic or Elastic-Viscoplastic material models;
- Support for complex geometry.

These requirements constitute the minimum attributes required for an algorithm to be applied to the FSW process analysis [41].

A 3-D numerical simulation of FSW concerned to study the impact of tool moving speed in relation to heat distribution as well as residual stress is presented by Riahi and Nazari [10]. Another interesting study presents a 3-D thermomechanical model of FSW based on CFD analysis [14]. The model describes the material flow around the tool during the welding operation. The base material for this study was an AA2024 sheet with 3.2 mm of thickness. The maximum and minimum temperature values in the workpiece (close to the tool shoulder) are shown in Fig. 11, where we can see that the maximum temperature value decreases when the welding velocity increases. In the other hand the maximum temperature value increases when the tool rotational velocity increases.

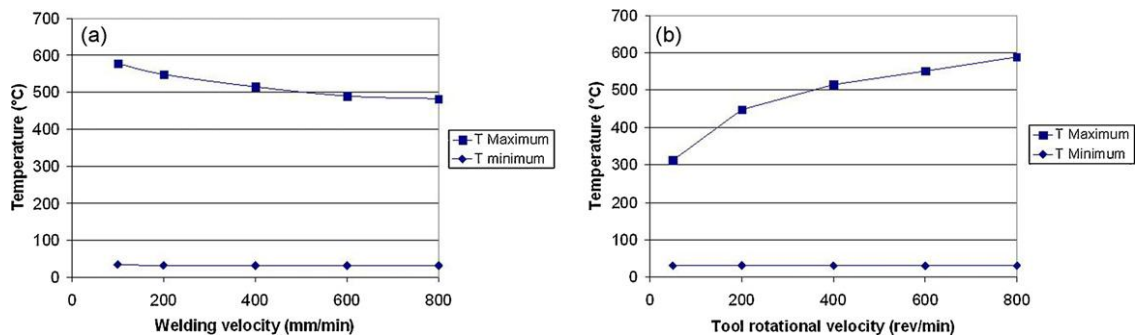

Fig. 11 Extreme temperatures in the welds; (a) as a function of the welding velocity of the tool for a tool rotational velocity equal to 400 rpm; (b) as a function of the tool rotational velocity for a welding velocity equal to 400 mm/min [14]

The model also provides data on the process power dissipation (plastic and surface dissipation contributions). The plastic power partition is made through the estimation of the sliding ratio in the contact between the tool and the workpiece. The predicted and measured evolution of the power consumed in the weld as a function of the welding parameters is presented in Fig. 12. Fig. 12 (a) shows the repartition of the predicted power dissipation as a function of the welding velocity. It is possible to see that although the total power generated in the weld increases with the welding velocity, the maximum temperature value decreases.

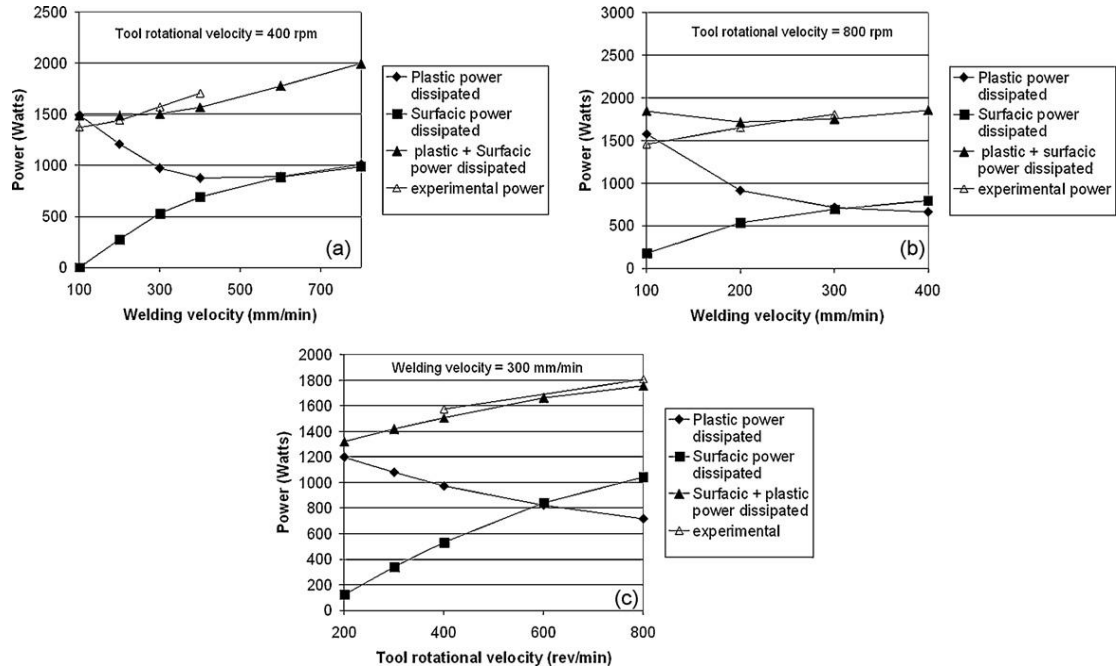

Fig. 12 Repartition of the predicted power dissipation in the weld between plastic power and surface power as a function of: (a) and (b) the welding velocity; (c) the tool rotational velocity [14]

## 6. CONCLUSIONS

FSW modeling helps to visualize the fundamental behavior of the welded materials and allows to analyze the influence of different weld parameters (including tool design) and boundary conditions, without performing costly experiments. FSW modeling is challenging task due to its multiphysics characteristics. The process combines heat flow, plastic deformation at high temperature, and microstructure and property evolution. Thus, nowadays, the numerical simulation of FSW process still cannot be used to optimize the process. The increasing knowledge produced about the process and computer resources can lead, maybe in a near future, to the use of numerical simulation of FSW to predict a good combination of the process parameters, replacing the experimental trials actually used. This will help to promote and expand the FSW process to a wider range of different applications and users.